\documentclass[%
 aip,
 amsmath,amssymb,
 reprint,%
]{revtex4-1}

\usepackage{graphicx}
\usepackage{dcolumn}
\usepackage{bm}

\usepackage[utf8]{inputenc}
\usepackage[T1]{fontenc}
\usepackage{mathptmx}
\usepackage{etoolbox}

\makeatletter
\def\@email#1#2{%
 \endgroup
 \patchcmd{\titleblock@produce}
  {\frontmatter@RRAPformat}
  {\frontmatter@RRAPformat{\produce@RRAP{*#1\href{mailto:#2}{#2}}}\frontmatter@RRAPformat}
  {}{}
}%
\makeatother
\begin{document}

\preprint{AIP/123-QED}

\title{Effects of gravity induced pressure variations for thermal liquid-gas phase-change simulations with the pseudopotential lattice Boltzmann method}
\author{Luiz Eduardo Czelusniak}
\affiliation{Center for Energy and Petroleum Studies, University of Campinas.}
 
\author{Luben Cabezas G{\'o}mez}%
\affiliation{Department of Mechanical Engineering, Engineering School of S{\~a}o Carlos, University of S{\~a}o Paulo.}

\author{Alexander J. Wagner}
\affiliation{
Department of Physics, North Dakota State University.}
\email{alexander.wagner@ndsu.edu.}
\date{\today}

\begin{abstract}
Direct simulations of phase-change and phase-ordering phenomena are becoming more common. Recently qualitative simulations of boiling phenomena have been undertaken by a large number of research groups. One seldom discussed limitations is that large values of gravitational forcing are required to simulate the detachment and rising of bubbles formed at a bottom surface. The forces are typically so large that neglecting the effects of varying pressure in the system becomes questionable.
In this paper we examine the effect of large pressure variations induced by gravity using pseudopotential lattice Boltzmann simulations. These pressure variations lead to height dependent conditions for phase co-existence and nucleation of either gas or liquid domains. Because these effects have not previously been studied in the context of these simulation methods we focus here on the phase-stability in a one dimensional system, rather than the additional complexity of bubble or droplet dynamics. Even in this simple case we find that the different forms of gravitational forces employed in the literature lead to qualitatively different phenomena, leading to the conclusion that the effects of gravity induced pressure variations on phase-change phenomena should be very carefully considered when trying to advance boiling and cavitation as well as liquefaction simulations to become quantitative tools.
\end{abstract}

\maketitle

%




\section{\label{sec:intro}Introduction}

In many industrial applications, such as nuclear reactors, electronic cooling, and refrigeration and power systems, boiling is a key phenomenon for heat transfer because the latent heat involved in the phase-change is significant \cite{carey2020liquid}. The boiling process involves the formation, growth and departure of bubbles or of a vapor layer depending on the boiling regime. In all cases the gravity plays a significant role acting through the buoyancy force and helping to detach the vapor structures from the heated surface. Jiang \textit{et al.} \cite{jiang2013dynamic} demonstrated that in a pool boiling bubble detachment process the buoyancy force is dominant for Jacob numbers equal and above 90. For small Jacob numbers, of about 10, the forces due to gravity and surface tension are the two dominant, being of the same magnitude. These findings indicate the importance of gravity force in boiling. Due to the efficiency of boiling process in transferring heat, there is a great interest in the use of numerical methods to examine this phenomenon. Lattice Boltzmann methods (LBM) have established themselves as excellent method candidates for simulating such boiling processes \cite{mu2017nucleate,sun2013three}. One of the most popular extensions of LBM for multiphase problems is the pseudopotential method \cite{chen2014critical,li2016lattice}.

In the pseudopotential literature, there are different approaches adopted by the authors to incorporate the effects of gravitational force in the simulations. Some authors use a conventional formulation for this force as $\bm{F}_g=\rho\bm{g}$ \cite{kang2002displacement,kang2005displacement,mazloomi2013thin,sankaranarayanan2002analysis,czelusniak2020force}. Other authors opt for a formulation in the form $\bm{F}_g=(\rho-\langle\rho\rangle)\bm{g}$, where $\langle\rho\rangle$ is the average density of the domain \cite{li2015lattice,gong2015lattice,fang2017lattice,sun2017numerical,chang2019lattice,wang2020mesoscale,sym12081358,JARAMILLO2022118705,ezzatneshan2023study}. If the force of gravity is small enough to lead to only minor differences in the pressure in the simulation, those formulations will give very similar results. In the literature, however, almost universally an unphysically large magnitude of the gravitational force is employed to accelerate the movement of bubbles. The corresponding effect on the thermodynamics of the system, however, has received scant attention. It will be shown in this work that for the amount of gravity usually employed the different formulations can lead to very different results. This behavior is related to the pressure field induced in the domain. The phase change is determined by the equation of state (EOS) relating pressure, temperature and density. When a condition of thermodynamic instability is reached, the equation of state induces the formation of a new phase in the system \cite{li2020does}. The magnitude of the pressures reached can be crucial for this phenomenon.

At first it may seem that the formulation choice is something of little relevance, because in many systems it is expected that the pressure variations in the domain due to gravity are negligible and thus the system will behave as if it was at a constant pressure.
We estimate how significant the effect of gravity is on simulations found in the pseudopotential method literature. The authors generally simulate a system that contains a column of liquid with a heated surface underneath where vapor bubbles form. We compare the pressure variation due to the presence of the liquid column with the critical pressure of the fluid, ie $\Delta p_l/p_c=\rho_l\cdot g\cdot h_l/p_c$ (where $\rho_l$ is the liquid density, $\Delta p_l$ is the liquid pressure variation, $h_l$ is the liquid column height and $p_c$ is the critical pressure).
In the case where the formulation $\bm{F}_g=(\rho-\langle\rho\rangle)\bm{g}$ is used
the ratio between the pressure variation in the liquid column and the critical pressure can be calculated by $\Delta p_l/p_c=(\rho_l-\langle\rho\rangle)\cdot g\cdot h_l/p_c$.

Obviously for a thermal problem due to temperature gradients the liquid density $\rho_l$ varies along the domain. However, we can do this calculation for the initial system (which is at constant temperature) used by several authors such as \citeauthor{li2015lattice} \cite{li2015lattice}, \citeauthor{fei2020mesoscopic} \cite{fei2020mesoscopic} and \citeauthor{feng2021investigation} \cite{feng2021investigation}. For these works we obtain ratios of $\Delta p_l/p_c$ of 7.4\%, 22.2\% and 21.5\%, respectively.
These are considerable pressure variations. As a real example, in a pot with a water  column of height $10$ $cm$ (at room temperature), this proportion is approximately 0.0045\%. In order to obtain a 10\% of variation for water (at room temperature and considering constant density measured at atmospheric pressure) it would be necessary a liquid column of approximately 220 $m$. Which means, that in most real situations, pressure variation would be completely negligible in comparison to the critical pressure.

We consider that there are two main reasons that influence authors to simulate a system where gravity has a significant effect. The first is related to the bubble detachment time. Normally, after the formation of an initial vapor nucleus, a bubble needs to increase its diameter several orders of magnitude before detachment occurs \cite{gao2022experimental}. This type of simulation would require the definition of a very large computational domain and would have a very high computational cost. To accelerate the detachment of the bubbles, the authors may be increasing the impact of the gravitational force in the simulations. This causes the bubbles to detach with smaller diameters \cite{gong2013lattice}.

A possible second reason why the authors simulate systems where gravity is more significant is related to the Mach number of the simulations. As the typical velocities of many systems are much lower than the real sound speed of ``incompressible" fluids such as water, numerical simulations become very expensive. Thus, the Mach number is relaxed, in order to allow simulations that behave in an approximately incompressible way but that have an adequate computational cost. In the LBM this can be done by adjusting the parameters of the equation of state (EOS) to reduce the speed of sound in the fluid. For the Carnahan-Starling (C-S) EOS the speed of sound at constant temperature is related to critical pressure as follows $v_s^2=(\partial p_{EOS}/\partial \rho)_T=p_c/\rho_c(\partial p_r/ \partial \rho_r)_{T_r}$ (with $\rho_r$, $p_r$ and $T_r$ the reduced properties).
For C-S EOS the term $(\partial p_r/\partial \rho_r)_{T_r}$ is constant given $\rho_r$ and $T_r$ and is independent of the choice of parameters that define the equation. So, one way to reduce the speed of sound is to reduce $p_c$, which makes the pressure variations due to gravity much more significant.

In this context where the gravity forces used produce high pressure variations compared to the critical pressure of the fluid, the phase change process will be affected. To better understand this influence of gravity, we analyzed a simple system composed of the vapor-liquid phases in which a theoretical equilibrium solution of this system could be obtained. Based on this theoretical solution, it was possible to make predictions of the conditions in which the system would present a phase change process based on the states of thermodynamic instability given by the equation of state. A good agreement was observed between the theoretical and numerical behavior obtained through simulations with the lattice Boltzmann method.

The initial motivation of the work was related to the boiling process, however it was observed that these large pressure variations are related to the most varied types of phase change processes such as vapor condensation due to high pressures or liquid cavitation due to low pressures. Thus, the study carried out considers the effect of gravity on phase change processes in general. As a main result, a significant difference was observed in the behavior obtained with the different formulations for gravitational force presented. The formulation based on the average density presented unphysical results for moderate and higher values of gravitational acceleration.

The article is organized as follows: In Sec.~\ref{sec:Gravity_Background}, a review is carried out regarding the gravitational forces used in LBM simulations of boiling, with emphasis on the pseudopotential method. The Sec.~\ref{sec:PseudopotentialLBM} summarizes the LBM and the pseudopotential approach for multiphase flows. In Sec.~\ref{sec:Thermodynamic_Study}, a thermodynamic study of a simple system consisting of a liquid and a vapor phase is carried out. A theoretical solution is developed for this system and compared with LBM numerical simulations. In Sec.~\ref{sec:Thermal_Conditions}, we analyzed under which conditions the defined system will present a phase change process. In Sec.~\ref{sec:Equilibrium_Configurations} the different configurations that the defined system can present in equilibrium are analyzed. Finally, in Sec.~\ref{sec:conclusion} a conclusion to the work is presented.




\section{\label{sec:Gravity_Background}Gravitational force in the pseudopotential method literature}

According to \citeauthor{chen2014critical} \cite{chen2014critical}, there are three common approaches of introducing the gravitational force in the LBM multiphase method.
The first one, is the physically correct form:
\begin{equation}
\label{eq:Gravity_1}
    \bm{F}_g = \rho \bm{g}.
\end{equation}
Another, more popular approach to introduce gravity considers the difference between the density to the average density $\langle \rho \rangle$:
\begin{equation}
\label{eq:Gravity_2}
    \bm{F}_g = \left( \rho - \langle \rho \rangle \right) \bm{g}.
\end{equation}
This approach introduces a second constant force that is independent of density and partially counteracts the force of gravity.
In addition, there are some works that used the following form:
\begin{equation}
\label{eq:Gravity_3}
    \bm{F}_g = \left( \rho - \rho_v \right) \bm{g},
\end{equation}
where $\rho_v$ is the vapor density. This force only affects the liquid phase. Although the first form, Eq.~(\ref{eq:Gravity_1}), is the one found in any physics textbook,
and it is employed in several pseudopotential LBM works \cite{kang2002displacement,kang2005displacement,mazloomi2013thin,sankaranarayanan2002analysis}, it is not the preferred method in boiling simulations as it will be discussed later.

The second form of the gravity force, Eq.~(\ref{eq:Gravity_2}), is the only form that allows the simulation of a rising bubble in a periodic simulation box because the net force on the fluid is guaranteed to be zero.
The second gravity form was used by \citeauthor{sankaranarayanan1999bubble} \cite{sankaranarayanan1999bubble} and \citeauthor{wagner2000simulation} \cite{wagner2000simulation}
to simulate a bubble rising in a liquid domain with periodic boundary conditions.
Some LBM papers refer to the work of \citeauthor{bunner2002dynamics} \cite{bunner2002dynamics} when using this gravity approach. The authors of this paper studied the motion of bubbles in periodic domains using a finite-difference discretization of the Navier-Stokes equations. 
Over time, the pseudopotential method was applied to more complex problems such as pool boiling or boiling inside channel flows. For these applications, the second form of gravity became widely used \cite{li2015lattice,gong2015lattice,fang2017lattice,sun2017numerical,chang2019lattice,wang2020mesoscale,ezzatneshan2023study}.

The third gravity formulation, given by Eq~(\ref{eq:Gravity_3}), can be found in the work of \citeauthor{markus2011simulation} \cite{markus2011simulation}, where the authors used the pseudopotential method. The same gravity form is found in the work of \citeauthor{takada2000numerical} \cite{takada2000numerical} with the binary fluid model developed by \citeauthor{swift1996lattice} \cite{swift1996lattice}. 

\citeauthor{amaya2010single} \cite{amaya2010single} compared the three gravitational forces in a bubble rising problem solved by a Cahn-Hilliard based LBM multiphase model. It was used a confined domain with no-slip boundary conditions in all directions. The difference between the terminal Reynolds number was less than 1\% in the test cases. The conclusion was that the particular form of gravity has little influence in the simulation outcome. 
\citeauthor{chen2014critical} \cite{chen2014critical} comment that numerical simulations of a bubble rising in a confined domain indicate that the way of calculating the gravitational force affects the numerical stability, and that Eq.~(\ref{eq:Gravity_2}) performs best for multiphase flow with high density ratios. 

For liquid-gas systems, as are relevant in boiling simulations, however, there are additional considerations that have so far been neglected in the literature.
In this paper we examine the impact of different version of the forcing term on boiling simulations specifically but also any general liquid-gas simulation in the presence of gravity. We show that the differences can be significant for the simple reason that each formulation induces a different pressure field in the simulations. In single phase problems, only the gradient of the pressure is important. But, for multiphase simulations, the equation of state relates the pressure with density and temperature. The equation of state is responsible for induce phase separation when the fluid reaches a state of thermodynamic non-equilibrium. Therefore, not only the gradient of the pressure is important, but also the actual value of the pressure. For conciseness only the first and second formulations of gravity which are the most employed ones in the pseudopotential literature will be analyzed in this work.




\section{\label{sec:PseudopotentialLBM}Pseudopotential lattice Boltzmann method}

In this section, we describe the basic equations of the pseudopotential LBM employed in the simulations presented here. It was used the same formulation of \citeauthor{li2015lattice} \cite{li2015lattice}. For more details there are some reviews about the pseudopotential LBM in the literature \cite{chen2014critical,li2016lattice}.

The lattice Boltzmann equation (LBE) can be written as
\begin{equation} 
\label{eq:LBE}
f_i(t+1,\bm{x}+\bm{c}_i) - f_i(t,\bm{x}) = \Omega_i(\bm{f},\bm{f}^{eq}) + F_i',
\end{equation}		
where $f_i$ are the particle distribution functions related to the particle velocity $\bm{c}_i$ and $f_i^{eq}$ are the local equilibrium distribution functions. $\bm{f}$ and $\bm{f}^{eq}$ correspond to vectors whose components are $[\bm{f}]_i=f_i$ and $[\bm{f}]^{eq}_i=f^{eq}_i$, and variables $t$ and $\bm{x}$ are time and space coordinates, respectively. 
The term $\Omega_i(\bm{f},\bm{f}^{eq})$ is the collision operator, in general dependent on $\bm{f}$ and $\bm{f}^{eq}$. 
The multiple-relaxation time (MRT) collision operator is given by
\begin{equation} 
\label{eq:OperatorMRT}
\Omega_i(\bm{f},\bm{f}^{eq}) = - \left[ \bm{M}^{-1} \bm{\Lambda} \bm{M} \right]_{ij}  (f_j - f_j^{eq}),
\end{equation}	
where $\bm{M}$ is the matrix that converts $(\bm{f} - \bm{f}^{eq})$ into a set of physical moments.
The velocity set used is the regular two-dimensional nine velocities one (D2Q9):
\begin{equation} 
\label{eq:VelocitySet}
\bm{c}_i =
\begin{cases} 
      (0,0), ~~~~~~~~~~~~~~~~~~~~~~~~~~~~~~~~~~ i = 0,  \\
      (1,0), (0,1), (-1,0), (0,-1), ~~~~~ i = 1,...,4, \\
      (1,1), (-1,1), (-1,-1), (1,-1), ~ i = 5,...,8. \\
   \end{cases}
\end{equation}		

The specific form of $\bm{M}$ used in this work was described in detail in \citeauthor{lallemand2000theory} \cite{lallemand2000theory} and it is obtained by a Gram-Schmidt procedure \cite{kruger2017lattice}:
\begin{equation} 
\label{eq:MatrixMRT}
\bm{M} = 
 	\begin{pmatrix}
 		1 & 1 & 1 & 1 & 1 & 1 & 1 & 1 & 1  \\
 		-4 & -1 & -1 & -1 & -1 & 2 & 2 & 2 & 2 \\ 
 		4 & -2 & -2 & -2 & -2 & 1 & 1 & 1 & 1 \\
 		0 & 1 & 0 & -1 & 0 & 1 & -1 & -1 & 1 \\
		0 & -2 & 0 & 2 & 0 & 1 & -1 & -1 & 1 \\
 		0 & 0 & 1 & 0 & -1 & 1 & 1 & -1 & -1 \\
 		0 & 0 & -2 & 0 & 2 & 1 & 1 & -1 & -1 \\
 		0 & 1 & -1 & 1 & -1 & 0 & 0 & 0 & 0 \\
 		0 & 0 & 0 & 0 & 0 & 1 & -1 & 1 & -1 \\
 		\end{pmatrix}.
\end{equation}
This form of matrix $\bm{M}$ was chosen such that collision matrix $\bm{\Lambda}$ responsible for the relaxation to local equilibrium would become diagonal in the moment basis. $\bm{\Lambda}$ can be written as
\begin{equation} 
\label{eq:RelaxationMatrix}
\bm{\Lambda} = \text{diag}\left( \tau_{\rho}^{-1},\tau_{e}^{-1},\tau_{\varsigma}^{-1},
\tau_{j}^{-1},\tau_{q}^{-1},\tau_{j}^{-1},\tau_{q}^{-1},
\tau_{\nu}^{-1},\tau_{\nu}^{-1} \right).
\end{equation}
where the parameters $\tau$ are the different relaxation times for the physical moments. The subscript of the relaxation times indicates the physical meaning of the moments. Of particular note are the conserved mass moment $\rho$ and the $x-$ and $y-$currents $j$, where the values of the relaxation times are arbitrary, since the moments do not change in collisions. The moments related to stress tensor trace $e$ control the bulk viscosity, while those related to the remainder of the stress tensor $\nu$ control the shear viscosity. The three remaining moments, related to symbols $\varsigma$ and $q$, are associated with kinetic moments and freely adjustable towards improving the stability of the method.

The last term on the right-hand side of Eq.~(\ref{eq:LBE}), $F_i'$, 
defines the forcing scheme, 
i.e. it adds the effects of an external force field, $F_{\alpha}$, 
to the macroscopic conservation equations. 
The external force is composed by the molecular interaction force and the gravitational force $\bm{F}=\bm{F}_{\text{int}}+\bm{F}_g$. 
In this work we use the forcing scheme defined by \citeauthor{li2015lattice} \cite{li2015lattice}, where $[\bm{F}']_i=F'_i$:
\begin{equation}
\label{eq:LiFS}
    \bigg( \bm{I} - \frac{\bm{\Lambda}}{2} \bigg)^{-1} \bm{M} \bm{F}' = 
 	\begin{pmatrix}
 		0 \\
 		6 \left( u_x F_x + u_y F_y \right) 
 		+ \frac{\sigma|\bm{F}_{\text{int}}|^2}{\psi^2 (\tau_e-0.5)}\\ 
 		- 6 \left( u_x F_x + u_y F_y \right)
 		- \frac{\sigma|\bm{F}_{\text{int}}|^2}{\psi^2 (\tau_{\varsigma}-0.5)}\\
 		F_x \\
		-F_x \\
 		F_y \\
 		-F_y \\
 		2 \left( u_x F_x - u_y F_y \right) \\
 		u_x F_y + u_y F_x  \\
 		\end{pmatrix},
\end{equation}
where $\bm{I}$ is the identity matrix. 
The parameter $\sigma$ is used to control the coexistence densities, for more details refer to the work of \citeauthor{li2013lattice} \cite{li2013lattice} and \citeauthor{li2015lattice} \cite{li2015lattice}. 

The relation between particle distribution functions $f_i$ and actual fluid velocity $\bm{u}$ depends on the forcing scheme.
In this forcing scheme, density and velocity 
fields are given by

\begin{subequations}
\begin{equation} 
\label{eq:Density}
\rho = \sum_i f_i,
\end{equation}	
\begin{equation} 
\label{eq:Momentum}
\rho \bm{u} = \sum_i f_i \bm{c}_i + \frac{\bm{F}}{2}.
\end{equation}	
\end{subequations}

A popular form of the equilibrium distribution function $f_i^{eq}$ is \cite{kruger2017lattice}:
\begin{equation} 
\label{eq:EquilibriumDistribution}
f_i^{eq} = w_i \bigg( \rho + \frac{c_{i \alpha}}{c_s^2} \rho u_{\alpha}
+ \frac{(c_{i \alpha}c_{i \beta} - c_s^2 \delta_{\alpha \beta})}{2 c_s^4} 
\rho u_{\alpha}u_{\beta} \bigg),
\end{equation}	
where terms $w_i$ are the weights related to each velocity $\bm{c}_i$, and $c_s$ is
the lattice sound speed. Note this definition differs from the original one of Qian \textit{et al.} \cite{qian1992lattice} because here the velocity in the equilibrium distribution contains the forcing correction of Eq. (\ref{eq:Momentum}).
The weights given by $w_i$
are $w_0 = 4/9$, $w_{1,2,3,4} = 1/9$ and $w_{5,6,7,8} = 1/36$, and the sound speed $c_s$ can take only the value $1/\sqrt{3}$.

Shan and Chen \cite{shan1993lattice} proposed an interaction force based on nearest-neighbor interactions (see Shan \cite{shan2008pressure} for the definition of nearest-neighbor interactions):
\begin{equation}
\label{eq:ShanChenForce}
    F_\alpha^{SC} = - G \psi  ( \mathbf{x} )
\sum_i w ( | \mathbf{c}_i |^2 ) \psi ( \mathbf{x} + \mathbf{c}_i ) c_{i\alpha},
\end{equation}
where $\psi$ is a density-dependent interaction potential and $G$ is a parameter that controls the strength of interaction. Weights $w( | \bm{c}_i |^2 )$ are $w(1)=1/3$ and $w(2)=1/12$.
We used the same definition of interaction potential as proposed by \citet{yuan2006equations} to add arbitrary equations of state $p_{EOS}$ to the system:
\begin{equation}
\label{eq:InteractionPotential}
    \psi(\rho) = \sqrt{ \frac{2\left( p_{EOS} - \rho c_s^2 \right)}{G c^2} }.
\end{equation}
When this technique is used, parameter $G$ no longer controls the interaction strength.

The LBE describes the particle distribution function evolution, however, the variables 
of interest are the macroscopic flow fields. Using the third-order analysis proposed by \citeauthor{lycett2015improved} \cite{lycett2015improved} we obtain the following mass and momentum conservation equations:

\begin{subequations}
\begin{equation} 
\label{eq:MassConservation}
\partial_t \rho + \partial_{\alpha} (\rho u_{\alpha}) = 0,
\end{equation}	
\begin{equation} 
\label{eq:MomentumConservation}
\partial_t (\rho u_{\alpha}) + \partial_{\beta} (\rho u_{\alpha} u_{\beta}) =
- \partial_{\beta} p_{\alpha\beta}
+ \partial_{\beta} \sigma_{\alpha\beta}' + F_{\alpha},
\end{equation}	
\end{subequations}
where the viscous stress tensor, $\sigma_{\alpha\beta}'$, can be written as \cite{kruger2017lattice}
\begin{equation}
\label{eq:StressTensor}
    \sigma_{\alpha\beta}' = 
    \rho \nu \left( \partial_{\beta} u_{\alpha} + \partial_{\alpha} u_{\beta} 
    - \frac{2}{3}\delta_{\alpha\beta}\partial_{\gamma}u_{\gamma} \right) 
    + \rho \nu_B \delta_{\alpha \beta} \partial_{\gamma}u_{\gamma},
\end{equation}
and kinematic viscosities $\nu$ and $\nu_B$ are related to the relaxation times of LBM through
\begin{equation}
\label{eq:KinematicViscosity}
    \nu = c_s^2 \left( \tau_{\nu} - 0.5 \right), ~~~~~ 
    \nu_B = c_s^2 \left( \tau_{e} - 0.5 \right) - \frac{\nu}{3}.
\end{equation}

The pressure tensor $p_{\alpha\beta}$ is given by: 
\begin{align}
    p_{\alpha\beta} = \bigg( &
    p_{EOS} 
    + \sigma \frac{G^2c^4}{6} ( \partial_{\gamma} \psi )( \partial_{\gamma} \psi )
    \nonumber\\&\left.
    + \frac{Gc^4}{12} \psi \partial_{\gamma} \partial_{\gamma} \psi 
    \right) \delta_{\alpha\beta} 
    + \frac{Gc^4}{6} \psi \partial_{\alpha} \partial_{\beta} \psi.
\label{eq:GeneralPressureTensor}
\end{align}

For the hydrodynamic boundary condition in the solid walls it was used the \citeauthor{zou1997pressure} non-slip condition \cite{zou1997pressure}.
The interaction force (Shan-Chen force) at the boundaries
were set equal to zero.

The energy conservation equation, in terms of temperature, is the same used by \citeauthor{hazi2009bubble} \cite{hazi2009bubble}: 
\begin{equation}
\label{eq:Energy_Equation}
    \partial_t T = - u_{\gamma} \partial_{\gamma}T 
    + \frac{1}{\rho c_v} \partial_{\gamma}(\kappa\partial_{\gamma}T)
    -\frac{T}{\rho c_v} 
    \left( \frac{\partial p_{EOS}}{\partial T} \right)_{\rho} \partial_{\gamma}u_{\gamma},
\end{equation}
where $c_v$ is the specific heat capacity and $\kappa$ is the thermal conductivity.
For the spatial discretization we use the same second-order isotropic finite differences scheme employed in the works \citeauthor{lee2005stable} \cite{lee2005stable} and \citeauthor{li2015lattice} \cite{li2015lattice}. For the temporal discretization we use a second order Runge-Kutta scheme as described in details by \citeauthor{li2016lattice} \cite{li2016lattice}.




\section{\label{sec:Thermodynamic_Study}Minimal one dimensional system}

\begin{figure}
\centering
	\includegraphics[width=\columnwidth]{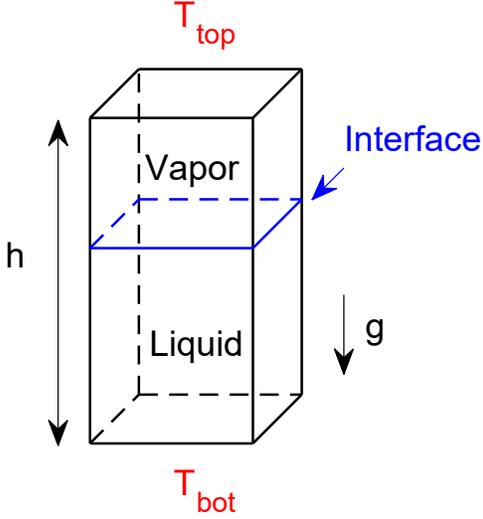}
	\caption{Scheme of the physical problem to be modelled.}
	\label{fig:Scheme_PhysSystem}
\end{figure}

Our goal is to understand how the gravitational field influences the nucleation process. It is instructive to focus on the simplest system consisting of an essentially one-dimensional situation shown in Fig.~(\ref{fig:Scheme_PhysSystem}). 
The system consists of two walls separated by a height $H$ where there are two phases separated by a planar interface, with liquid in the bottom and vapor in the top. Since it is an one-dimensional problem, there is no gradient of density, pressure and temperature in the transversal direction.
The bottom wall is maintained at a temperature $T_{\text{bot}}$ and the top wall is at $T_{\text{top}}$. The average density of the system is $\langle\rho\rangle$ and the magnitude of the gravitational acceleration is $g$. 
The proposed problem will be used as a case of study for the LBM numerical simulations along the work. Since we are working with a simple system, it was possible to obtain a theoretical solution for studying the gravity effect on nucleation process and for performing comparisons with LBM simulations.

\subsection{Theoretical solution}
\label{sec:Thermo_Profiles}

We will solve this problem theoretically by considering that the system is in equilibrium. This means that there will be no time-dependence and the fluid velocity $u_\alpha =0$ vanishes.  Since heat can flow only in y-direction (one-dimensional problem) and there is only heat sinks or sources at the boundary, the heat flux $q'$ is constant. 

In this simple case the continuity equation (\ref{eq:MassConservation}) is identically fulfilled, the Navier Stokes equation (\ref{eq:MomentumConservation}) becomes 
\begin{equation} \label{eq:Momentum_Balance_TS}
    \frac{dp(T,\rho)}{dy} = F_{g,y}(\rho).
\end{equation}
where $p=p_{yy}$ is a component (we will call as normal pressure) of the fluid pressure tensor of Eq.~(\ref{eq:GeneralPressureTensor}) and $F_{g,y}(\rho)$ is the vertical component of the gravitational force as a function of the local density.
For the energy conservation equation (\ref{eq:Energy_Equation}) we obtain
\begin{equation} \label{eq:Energy_TS}
    - \kappa \frac{dT}{dy} = q'.
\end{equation}
where $\kappa$ is the thermal conductivity.

We are looking for solutions $\rho(y)$ and $T(y)$ that fulfill Eqs.~(\ref{eq:Momentum_Balance_TS}) and (\ref{eq:Energy_TS}). This is a complex problem that can be simplified by modeling the interfacial regions between different phases as sharp interfaces that connect a liquid and a gas equilibrium densities corresponding to the temperature at the location of the interface. This allows us to invert the equation of state to give us the density as a function of the local pressure:
\begin{equation} \label{eq:Inverse_EOS}
    \rho = \rho_{EOS}(p,T).
\end{equation}

\begin{figure}
\centering
	\includegraphics[width=\columnwidth]{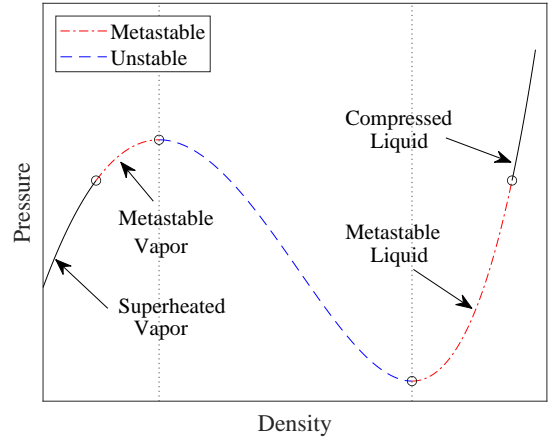}
	\caption{Illustration of the pressure-density relation given by an EOS for a non-ideal gas at a fixed temperature.}
	\label{fig:EOS}
\end{figure}
The function $\rho_{EOS}(p,T)$ is multivalued, as can be seen in Fig.~(\ref{fig:EOS}). The possible solutions we are looking for are on the stable and metastable branches. The solutions of the unstable branches are not of interest here. When switching between the different branches, those transitions correspond to an equilibrium interface and must therefore always connect equilibrium vapor densities to equilibrium liquid densities. 

In the simplest case the heat conductivity is constant (which is assumed for the analysis in this paper), which allows us to integrate Eq.~(\ref{eq:Energy_TS}) to obtain
\begin{equation} \label{eq:Temp_Profile}
    T(y) = -\frac{q'}{\kappa} y  + T(0).
\end{equation}
A more general dependence of $\kappa(\rho,T)$ can be taken into account in principle, and require a simultaneous numerical solution of the heat and momentum equations. This more general procedure is detailed in Appendix~\ref{sec:Appendix_1}.
The assumption of constant $\kappa$ also allows us to find $q'$ in terms of the boundary positions and temperatures, if required. This then fully determines the temperature profile $T(y)$ leaving us to find the density profile $\rho(y)$ by integrating Eq.~(\ref{eq:Momentum_Balance_TS}). To do this we pre-suppose an interface location, which will give us the equilibrium liquid and gas densities corresponding to the local temperature, and from those we can numerically integrate Eq.~(\ref{eq:Momentum_Balance_TS}) along the two different density branches. If in this integration at any point reaches the endpoints of the metastable branches are reached, it follows that a stable solution no longer exists. Special care need to be taken for situations with two interfaces, as discussed below. Thus we find the solutions for this simple one-dimensional system in the sharp interface limit.

To obtain the interface pressure $p_i$ and phase densities ($\rho_{v,i}$ and $\rho_{l,i}$) at the chosen interface location we use the Maxwell rule:
\begin{subequations} \label{eq:MaxwellRule}

\begin{equation} \label{eq:Integral_MaxwellRule}
\int_{\rho_{v,i}}^{\rho_{l,i}} \left( p_{i} - p_{EOS}(\rho,T_i) \right) \frac{d \rho}{\rho^2} = 0,
\end{equation}
\begin{equation} \label{eq:Pressures_MaxwellRule}
    p_{i}=p_{EOS}(\rho_{v,i};T_i); ~~~ p_{i}=p_{EOS}(\rho_{l,i};T_i);
\end{equation}
\end{subequations}
Eqs.~(\ref{eq:Integral_MaxwellRule}) and Eq.~(\ref{eq:Pressures_MaxwellRule}) are equivalent to requiring the chemical potential and pressures of the two phases infinitesimally close to the interface to be identical.

Often we require to find the solution for some given total amount of material, since that is what is given in our lattice Boltzmann simulations. To adjust the theoretical solutions we can adapt the location of the interface until the desired average density is achieved. Alternatively the lattice Boltzmann simulations can be initialized with a average density given by the theoretical solution. 

\subsection{Comparison between theoretical solution and LBM}

In this work, our focus will be to understand the role of gravity. For this, the following parameters of the theoretical solution will be fixed in all cases analyzed in this article: $h=100$, $\langle\rho\rangle=1.3344$. There are also two temperatures that can be varied $T_{\text{bot}}$ and $T_{\text{top}}$. To reduce the number of degrees of freedom in our analysis, we will work with an average temperature $T_{\text{med}}=(T_{\text{bot}}+T_{\text{top}})/2$ and the temperature difference between the walls $\Delta T = T_{\text{bot}}-T_{\text{top}}$. We decided to set $T_{\text{med}}/T_c=0.75$ and analyze the relationship between gravity and $\Delta T$. This choice of $T_{\text{med}}$ was made because it allows numerical simulations to occur in a range of temperatures in which the method is stable and not so close to the critical point. 

The lattice Boltzmann solutions were performed using the wet-node approach with a mesh of $N_x=101$ nodes.
Since the lattice Boltzmann method dynamically solves the energy conservation equation, the values of $c_v$ and $\kappa$ need to be specified. We set $c_v=5$ and $\kappa=0.02$ in most simulations. In some cases the value of $\kappa$ has been reduced to 0.008 to increase the stability of the method.
Note that as the simulations involve systems in equilibrium, the specific value of $k$ does not change the results, since the temperature profile will always be linear.
In the collision operator, $\tau=1$ was used for all relaxation times.
To ensure that we obtain the correct Maxwell rule we chose  $\sigma=1.385$ in the forcing scheme of Eq.~(\ref{eq:LiFS}) \cite{li2015lattice}. 
In the tests conducted in this article (both theoretical and numerical) it was selected the Carnahan-Starling (C-S) as the EOS:
\begin{equation} \label{eq:EOS}
    p_{EOS}^{C-S} = \rho R T \frac{1+b\rho/4+(b\rho/4)^2-(b\rho/4)^3}{(1-b\rho/4)^3}  - a \rho^2.
\end{equation}
where the parameters $a$, $b$ and $R$ depend on the critical fluid properties: $a\approx3.8533p_c/\rho_c^2$, $b\approx0.5218/\rho_c$ and $R\approx2.7864p_c/(\rho_cT_c)$. Here, we will simply use $a=0.5$, $b=4$ and $R=1$ in lattice units, as these are values commonly used in the literature \cite{li2013lattice,czelusniak2022shaping}.

\begin{figure}
\centering
	\includegraphics[width=\columnwidth]{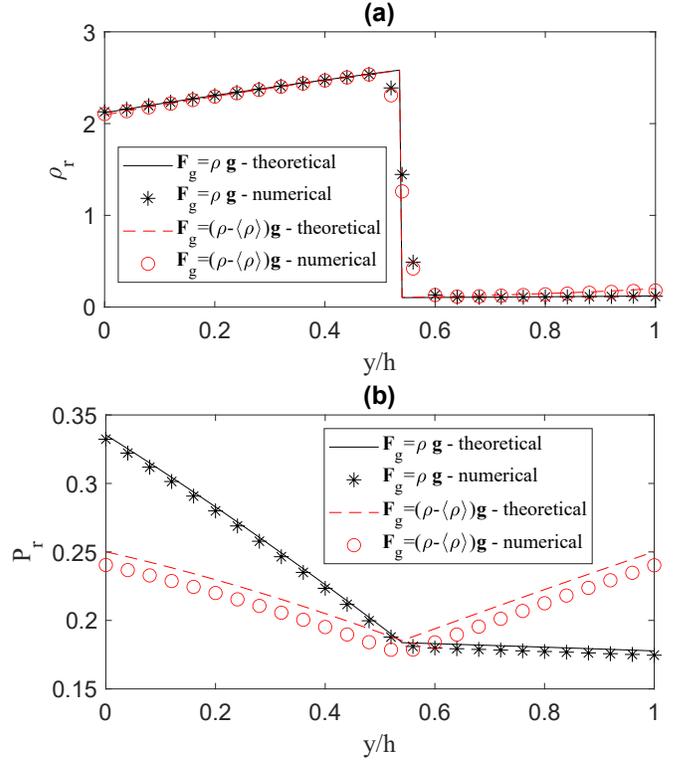}
	\caption{Theoretical and numerical (a) density and (b) normal pressure profiles for the example with $g=2\times10^{-5}$ and $\Delta T/T_c = 0.2$ using two gravity formulations, Eqs.~(\ref{eq:Gravity_1}) and (\ref{eq:Gravity_2}).}
	\label{fig:Example_Pressure_Profile}
\end{figure}

To illustrate the theoretical solution we first consider a system with a gravitational acceleration of  $g=2\times10^{-5}$, and a temperature difference $\Delta T/T_c = 0.2$. 
The theoretical density and pressure profiles computed using the gravity formulations of Eqs.~(\ref{eq:Gravity_1}) and (\ref{eq:Gravity_2}) are displayed in Fig.~(\ref{fig:Example_Pressure_Profile}). The solutions are presented in terms of reduced density $\rho_r=\rho/\rho_c$ and reduced pressure $p_r=p/p_c$. We can see that the density profiles are very similar, but the formulation of Eq.~(\ref{eq:Gravity_1}) produces a smaller pressure variation in the vapor region and a greater variation in the liquid region due to the difference of the phase densities. The formulation of Eq.~(\ref{eq:Gravity_2}) results in a sort of ``V'' shape pressure profile. 

We also compared the theoretical solutions with LBM simulations. 
The one-dimensional problem described above  will be modelled by a two-dimensional LBM simulation with periodic conditions in the side boundaries. 

The density profile shown in Fig.~(\ref{fig:Example_Pressure_Profile}.a) shows excellent agreement for both forms of gravity between the theoretical and lattice Boltzmann results, except in the interface that is diffuse in the lattice Boltzmann results by construction.

The LBM pressure results shown in Fig.~(\ref{fig:Example_Pressure_Profile}.b) were computed using the normal pressure $p_{yy}$ given by Eq.~(\ref{eq:GeneralPressureTensor}) with central finite differences to approximate the spatial derivatives. 
We can see in Fig.~(\ref{fig:Example_Pressure_Profile}.b) that for the gravity formulation of Eq.~(\ref{eq:Gravity_1}) the LBM results were very close to the theoretical solution for the pressure profile. For the formulation of Eq.~(\ref{eq:Gravity_2}) a deviation of about 5\% was observed between the LBM and theoretical solution for the pressure profile but apart from this constant offset the agreement is excellent.




\section{\label{sec:Thermal_Conditions}Thermodynamic conditions for new phase formation}

\begin{figure}
\centering
	\includegraphics[width=\columnwidth]{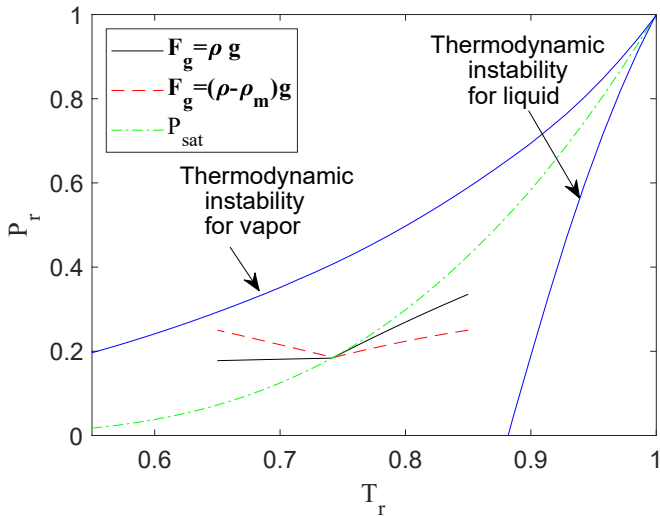}
	\caption{Pressure-temperature relation for the theoretical solution of the same example of Fig.~(\ref{fig:Example_Pressure_Profile}.b). The solid black line represents results for first gravity formulation and dashed red line represents the second formulation. The dash-dotted line represents the saturation pressure at each temperature. Solid lines with text indication represents regions of thermodynamic instability.}
	\label{fig:Example_Pressure_Temperature}
\end{figure}

The large difference between the observed pressures away from the interface for the two forms of gravity considered here suggests that these large deviations may strongly influence the formation of new phases at the boundaries. To better analyze this it is helpful to plot the results in a pressure/temperature graph.
Combining the results for the pressure shown in Fig.~(\ref{fig:Example_Pressure_Profile}.b) with the temperature of Eq.~(\ref{eq:Temp_Profile}) at each point we generate the pressure-temperature relations shown in Fig.~(\ref{fig:Example_Pressure_Temperature}).
This figure also contains the saturation pressure $p_{sat}(T)$ obtained from the Maxwell rule in green as well as lines of unconditional instability given by the spinodal points shown in Fig.~(\ref{fig:EOS}) in the dashed-blue line. The point where the system pressure profile coincides with the saturation line represents the liquid-vapor interface.

In the pressure-density curve (at a fixed temperature) of the EOS represented in Fig.~(\ref{fig:EOS}) there is a local maximum point and a local minimum. Between these two points there is a region of thermodynamic instability.
The instability lines added in Fig.~(\ref{fig:Example_Pressure_Temperature}) represent the pressures of these local maximum and minimum points given by the EOS at different temperatures. These pressures are denominated spinodal pressures.

Let us now consider the specific simulation results of Fig.~(\ref{fig:Example_Pressure_Profile}) and display them in the $P_r$-$T_r$ diagram of  Fig.~(\ref{fig:Example_Pressure_Temperature}).
The lower surface temperature of the system is $T_{\text{bot}}/T_c=0.85$ which represents the right extremities of the theoretical solutions (solid black line and dashed red line) in Fig.~(\ref{fig:Example_Pressure_Temperature}). For the two gravity formulations the thermodynamic state at this surface is far from the region of instability for the liquid. Thus both of the two formulations lead to stable solutions.

Unsurprisingly the two gravity formulations are notably distinct in Fig.~(\ref{fig:Example_Pressure_Temperature}). It is also easy to envisage the effect of a slight increase in gravity, which would lead to and increase of the slopes of the nearly straight lines. In the example shown here an increase in gravity would let the dashed line, corresponding to the second gravity formulation, Eq.~(\ref{eq:Gravity_2}), to contact the instability line of vapor. In that case we would predict the formation of a liquid phase at the top wall. Since an increase of gravity would not favour the formation of liquid at the top of the system in a real system, this is a spurious phenomenon caused by the unphysical gravity formulation.

\subsection{Example of phase formation}

In the previous section we used the regions of instability in the pressure relation to predict at what temperatures and pressures a transition from liquid to vapor or from vapor to liquid should be expected. This bulk argument is a little simple minded since nucleation phenomena are in general more complicated. Sometimes it is possible to derive analytical nucleation criteria \cite{foard2009enslaved} but here the presence of a wall with unexplored wetting properties poses a challenge. We briefly explored this nucleation phenomenon below and show that wall wetting phenomena can lead to a small shift of the instability points away from the ones predicted using purely bulk arguments.

In this section, we will simulate with LBM an example of a system that reaches the region of thermodynamic instability. We calculate that the system modeled by the first gravity formulation with the parameters $g=1.5\times10^{-5}$ and $\Delta T/T_c=0.3166$ (or equivalently $T_{\text{bot}}=0.9083$ and $T_{\text{top}}=0.5917$) must reach the region of instability for the liquid phase in the lower wall, and a new vapor phase (according to the assumptions used) must form at this location.

The idea with this test is to perform the simulation of a system with these boundary conditions that, in theory, would take it to a condition of thermodynamic instability. For this, the following simulation procedure was adopted. First, a system with boundary conditions close to the desired ones will be simulated, however, that still keep the system within the stability condition.
After this simulation (still stable) reaches equilibrium, the boundary conditions will be changed to the desired values. Thus, the system will evolve smoothly to the new condition and it will be possible to observe along the process if a new phase will form or if the system will be able to reach a new equilibrium without the phase change.

Thus, first a system was simulated with the upper wall temperature $T_{\text{top}}/T_c=0.6213$ instead of $T_{\text{top}}/T_c=0.5917$, a value 5\% higher. This higher temperature at the top wall increases the pressure of the entire system, preventing the phase change process from occurring in the first place.
After the system with these boundary conditions reaches the equilibrium, the upper wall temperature is reduced to the desired value and the pressure of the entire system begins to decrease, while the system evolves to the new equilibrium state.

\begin{figure}
\centering
	\includegraphics[width=\columnwidth]{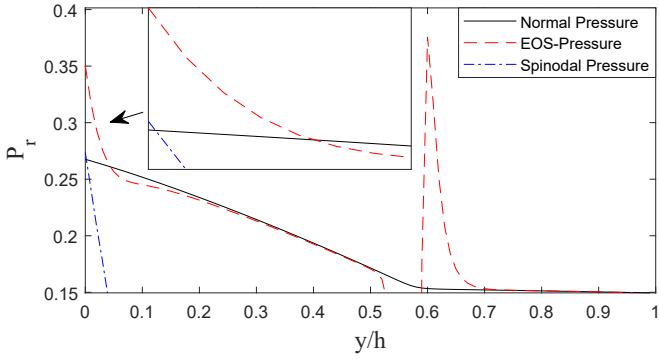}
	\caption{Equilibrium pressure profile for the system with: $g=1.5\times10^{-5}$ and $\Delta T/T_c=0.3166$. It was plotted the normal pressure and EOS-Pressure. Also the unstable pressure for the temperature of each point was displayed.}
	\label{fig:Predicted_Equi_Limit}
\end{figure}

In Fig.~(\ref{fig:Predicted_Equi_Limit}), we plotted the pressure profile of the system simulated with LBM in equilibrium. It was plotted the normal pressure, the pressure computed by the EOS and the spinodal pressure for the temperature of each point. We can see that in equilibrium, the normal pressure of this system touches the region of instability. However no phase change occurred. We used as a hypothesis that the EOS-pressure should be equal to the normal pressure out of the interface region. This is true in most of the liquid region, however, it has been observed that near the lower boundary there is a large deviation between these two pressures.
Thus, the fluid at that position is still not in the region of thermodynamic instability.

\begin{figure}
\centering
	\includegraphics[width=\columnwidth]{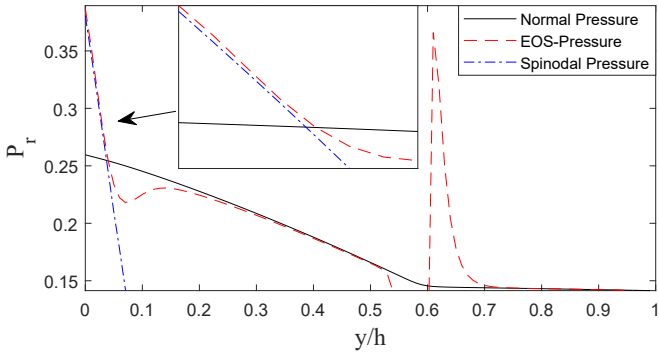}
	\caption{Equilibrium pressure profile for the system with: $g=1.5\times10^{-5}$ and $\Delta T/T_c=0.3381$.}
	\label{fig:True_Equi_Limit}
\end{figure}

In order to reach the region of thermodynamic instability another case was simulated increasing the temperature difference $\Delta T/T_c$. The simulation procedure is the same as the previous case, first a system was simulated with $T_{\text{top}}/T_c=0.6213$ and after equilibrium this temperature was reduced to the desired value.
When we increase $\Delta T/T_c$ by 6.79\% (from 0.3166 to 0.3381), we reach the limiting case where the system still exhibits no phase change. This limiting case is represented in Fig.~(\ref{fig:True_Equi_Limit}). We can see that the unstable pressure is much higher than the normal pressure and is almost coinciding with the EOS-pressure values.

The difference between the thermodynamic pressure and the normal pressure of the lattice Boltzmann method at the boundary is clearly a wall-induced effect that implements some specific wetting property of the wall. Since this is a small effect we decided that it is outside of the scope of our paper and the study of the details of this effect will be left to future research.

\begin{figure}
\centering
\includegraphics[width=\columnwidth]{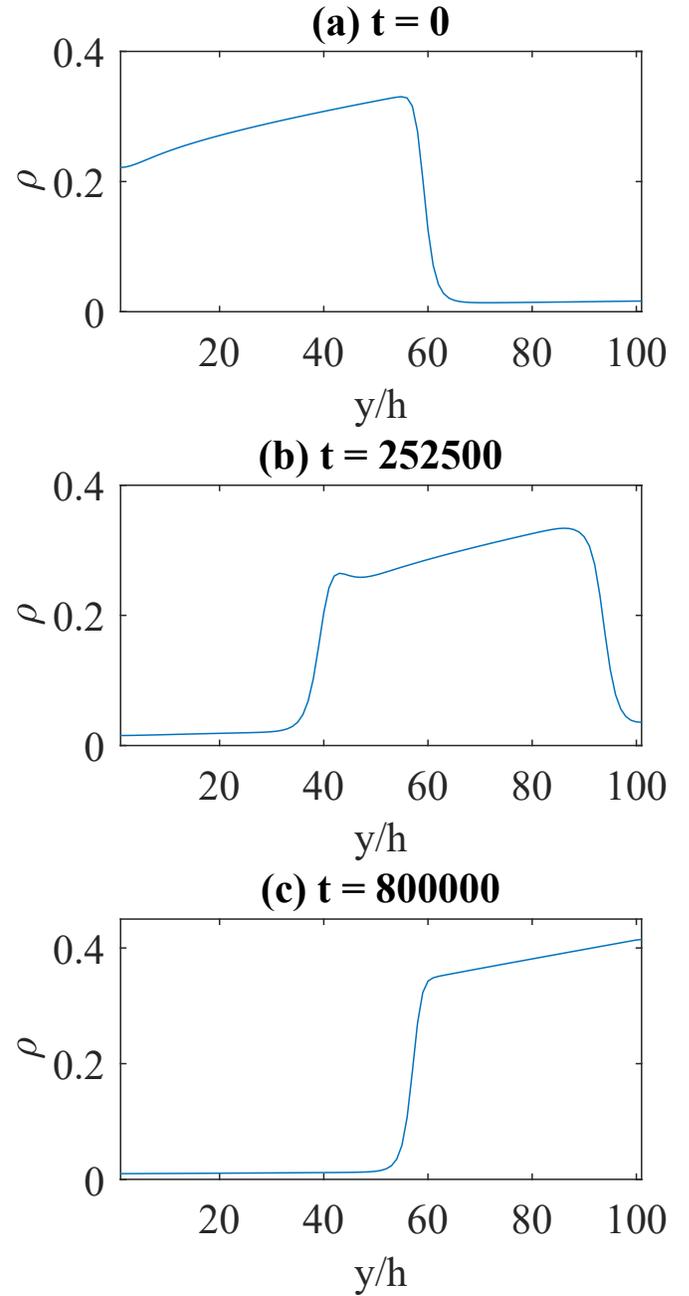}
	\caption{Density profiles of a system undergoing a phase change process. The system was defined with the parameters: $g=1.5\times10^{-5}$ and $\Delta T/T_c=0.3382$.}	\label{fig:Boiling_Densities}
\end{figure}

If we simulate the system with a higher temperature difference (following the same simulation procedure), a phase change process is observed. The density profile of such a system (with $\Delta T/T_c=0.3382$) at different time instants is shown in Fig.~(\ref{fig:Boiling_Densities}). At $t=0$, the initial condition is a system with liquid at the bottom and vapor at the top. Remember that the temperature on the top wall is $T_{\text{top}}/T_c=0.6213$ at this point to avoid phase change. Then, the upper wall temperature is reduced to the desired value and the system is allowed to evolve to equilibrium. The lower wall pressure decreases until the EOS-pressure matches the unstable pressure, and a new phase forms on the lower wall. At time $t=252500$ we can see a layer of vapor covering the lower surface. This vapor layer grows and at equilibrium at $t=800000$ the two phases have completely reversed and now there is vapor below and liquid above.

\subsection{Dependence of Boiling temperature on gravity}
\label{sec:Boiling_Temperature}

An important parameter for predicting the formation of a new phase is the magnitude of the gravitational acceleration.
For a stronger gravitational field, the pressure at the bottom wall will be higher, and a higher boiling temperature will be needed to observe the formation of a new phase.
Through the theoretical solution and simulations with LBM, we evaluated the impact of the first and second formulations of the gravitational force on the boiling temperature.

Various values of $g$ and $\Delta T$ were evaluated. In the tests performed, negative values of $g$ were also considered. This was done to try to emulate conditions similar to those that occur in a cavitation process. Inside pumps, a fluid can pass through a region of low pressure with the formation of vapor bubbles. In tests, it was verified what happens when the fluid pressure is reduced through the use of an artificial negative gravity.

Regarding the theoretical solution,
the maximum temperature difference $\Delta T$ that can be applied in a system (for a certain gravitational acceleration) without observing the formation of a new phase was computed.
For the simulations with the LBM, the following procedure was used. First, a system was simulated in a certain condition known to be stable in terms of gravity and temperature difference. Then, these parameters ($g$ and $\Delta T$) started to be slightly adjusted until reaching stability limits.
It was considered that a limiting case of stability is one in which a variation of only 0.5\% in $\Delta T$ or $g$ is sufficient for a phase change process to occur.

\begin{figure}
\centering
	\includegraphics[width=\columnwidth]{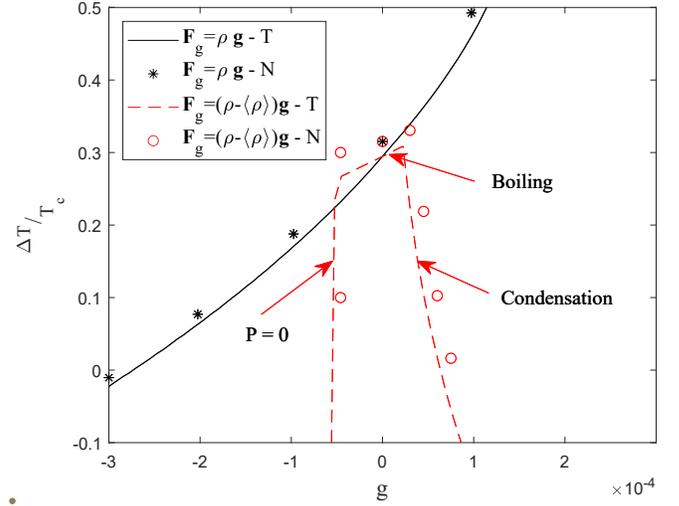}
	\caption{Temperature difference between walls as a function of the gravity acceleration. In the legend ``T" means theoretical results and ``N" means the LBM numerical results.}
	\label{fig:Boiling_Theoretical}
\end{figure}

Figure~(\ref{fig:Boiling_Theoretical}) shows the temperature difference required to induce the phase change as a function of gravity, both for the theoretical and numerical results and for the two gravity formulations.
First we compare the theoretical and numerical results. As previously mentioned, the deviation of the EOS-pressure from the normal pressure near the solid wall causes the numerical results to differ from the theoretical ones. However, when we look at Fig.~(\ref{fig:Boiling_Theoretical}) over a wide range of gravitational acceleration values, the behavior of the numerical is in good agreement with the theoretical curves.

A significant difference is observed when comparing the results with the two gravity formulations. First, we can see that for the second formulation, the curve can be divided into three regions with distinct behaviors as indicated in the curve. On the left side ($g<-4.5\times10^{-5}$), the curve looks almost like a vertical line and is indicated on the graph as ``$P=0$'', which means that the pressure of the vapor region at the top wall is equal to zero. This physical behaviour is explained as follows. 
For negative values of gravitational acceleration $g$, the vertical component of the gravitational force becomes $F_g=(\rho-\langle\rho\rangle)|g|$. As $\langle\rho\rangle>\rho$ in the vapor region, the force tends to decrease the pressure of the system towards the top wall. When the pressure reduces, the vapor density is also reduced. However, the gravitational force continues to act through the term $-\langle\rho\rangle|g|$ which makes the pressure of the system goes to zero.   
For zero pressure, the vapor density also goes to zero which makes the simulation unstable. This behavior is confirmed by numerical simulation. However for the simulation, the method becomes unstable before reaching zero pressure, and thus there is a small difference between the numerical and theoretical results. 

The second region is indicated as ``boiling'' on the chart ($-4.5\times10^{-5}\leq g\leq 2.25\times10^{-5}$). In this part of the curve, the phase change is due to boiling of the liquid at the bottom wall. We can see that the boiling temperature increases as the gravitational acceleration is increased.
For zero gravity the two formulations have the same maximum temperature difference for boiling. But the angle of the two curves is different. The force of gravity in the liquid region is smaller in the second formulation, because the density of the liquid is subtracted from the average density $\rho-\langle\rho\rangle$. Therefore, this formulation is less sensitive to changes in gravitational acceleration.

A third part of the curve ($g>2.25\times10^{-5}$) is indicated as ``condensation''. For the second gravity formulation, the pressure increases in the vapor region towards the top wall for positive $g$ as seen in Figure~(\ref{fig:Example_Pressure_Profile}).
Also, this pressure variation can be significant depending on the amount of liquid in the system. For some positive value of gravitational acceleration the pressure in the vapor region becomes high enough to induce condensation. This explains why there is a drastic change in the behavior of the phase change curve for the second formulation.

All this makes the second formulation differ greatly from the first (and from physical reality). For the first formulation, the pressure in the vapor region slightly varies due to the low density of the vapor compared to the density of the liquid. This small pressure variation in the vapor region compared to the liquid prevents the vapor from reaching zero pressure or the condensation condition, making the curve better behaved. Boiling is the only phase change mechanism in this range of gravitational accelerations and temperature differences.

For the first gravity formulation we can see that phase change happens even for $\Delta T=0$. In this case a negative gravity was used. This case is similar to the phenomenon of cavitation where the system pressure is reduced until the phase change process from liquid to vapor starts to occur.




\section{\label{sec:Equilibrium_Configurations}Equilibrium configurations}

For the simulations considered in this paper we focused on a liquid-gas coexistence with liquid at the bottom and vapor at the top. For situations where boiling at the bottom occurred we observed the formation of a second vapor phase at the bottom as shown in Fig.~(\ref{fig:Boiling_Densities}). Since we are only interested in one dimensional situations here, we don't observe the formation of bubbles, instead we found an inversion of the morphology where we then saw a vapor phase at the bottom of the simulation and a liquid at the top. 

This result suggests that there are sometimes multiple possible configurations of the morphology that are simultaneously stable, like a morphology with the liquid either at the top or at the bottom. It also raises the question whether there are other possible configurations, and we will see in this section that this is indeed the case. Apart from the two morphologies discussed above there is sometimes the possibility of having a liquid phase suspended in the middle of the vapor phase. 
We will give a theoretical analysis of the stability of those configurations below.
To simplify the presentation we will refer to the following configurations by acronyms: liquid below and vapor above (L-V); vapor underneath,
liquid in the middle and vapor on the top (V-L-V) and finally
vapor below and liquid above (V-L).

We want to examine existence of the possible morphology for given top and bottom temperatures and a given amount of material.
This check is done separately for each configuration. For the L-V system, the procedure is the same as in the previous sections: the location of the  interface $h_i$, in Equations~(\ref{eq:Interface_Vapor}) and (\ref{eq:Interface_Liquid}), is incrementally varied until a solution is found that has an average density equal to the desired value $\langle\rho\rangle$. If an equilibrium condition can not be found that satisfies this condition, this implies that this configuration cannot exist for the specified system.
For the V-L configuration, the calculation is very similar to that presented in Equations~(\ref{eq:Interface_Vapor}), (\ref{eq:Numerical_Vapor_Region}), (\ref{eq:Interface_Liquid}) and (\ref{eq:Numerical_Liquid_Region}), we just change the order of the indices $v$ and $l$.

For the V-L-V configuration the calculation is a little more complex. In this case, there are two vapor-liquid interfaces within the system and each interface is assumed to have the saturation pressure for the temperature at that location. This configuration is only possible if the pressure variation due to gravitational force along the liquid column is equal to the difference between the saturation pressures of the two interfaces.

Thus, the following solution procedure was adopted. First, we assume the location of the lower interface $h_i$ and solve the system as if the configuration were V-L. If, at some location in the liquid column, the fluid pressure is equal to the saturation pressure, it is assumed that there is a new interface at this point and it switches to a new vapor phase in the calculations. After that, the value of $h_i$ is incrementally varied until a possible case is found in which this system has an average density $\langle\rho\rangle$ equal to the desired value.

\begin{figure}
\centering
\includegraphics[width=\columnwidth]{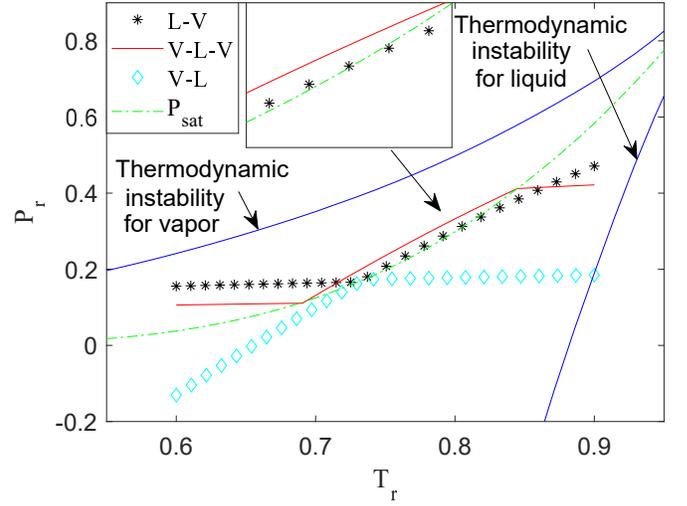}	\caption{Theoretical pressure-temperature relation for all possible phase configurations with the first gravity formulation, Eq.~(\ref{eq:Gravity_1}), and the parameters $g=4\times10^{-5}$ and $\Delta T/T_c=0.3$.}	\label{fig:Possible_Config}
\end{figure}

We show a specific example where all three configurations are possible (for the first gravity formulation) in Fig.~(\ref{fig:Possible_Config}) as a $P_r-T_r$ diagram. Any interface has to be located on the saturation pressure line. Because of the higher density of the liquid, the liquid branches of the graph have a higher slope. 

The L-V configuration is similar to that shown in Fig.~(\ref{fig:Example_Pressure_Profile}b) and Fig.~(\ref{fig:Example_Pressure_Temperature}). The high-temperature region corresponds to liquid at the bottom, which eventually meets the saturation pressure, where the interface to the vapor phase is found. The pressure in the vapor varies only slightly, because the weight of the vapor is much less than that of the liquid. This configuration becomes unstable when the liquid branch meets the blue line corresponding the instability for the liquid. When these branches meet boiling occurs at the top. Similarly the vapor branch becomes unstable when it meets the blue line for the instability of vapor, leading to temperature induced condensation at the top of the container. 

\begin{figure}
\centering
\includegraphics[width=\columnwidth]{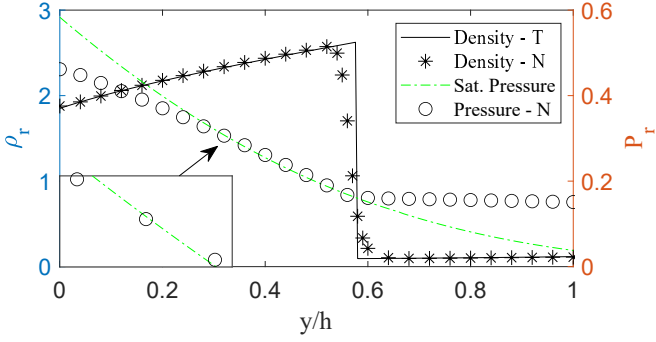}
	\caption{Density and pressure profile for the L-V example with $g=4\times10^{-5}$ and $\Delta T/T_c=0.3$. T means theoretical results and N numerical results. The numerical pressure showed here is the normal pressure.}	\label{fig:L_V_Comparison}
\end{figure}

Another feature of this configuration is shown in the inset of Fig.~(\ref{fig:Possible_Config}), where it can be seen that the liquid branch of the L-V morphology crosses the saturation pressure line. 
Until this moment, the system points that coincided with the saturation line were treated as interfaces. However, it is possible for a given phase of the fluid to exist stably in this condition because the fluid phase is still within the region of thermodynamic stability. This behaviour is shown in Fig.~(\ref{fig:L_V_Comparison}), where the pressure of this system and the density were plotted as functions of the position along the domain. In the Fig.~(\ref{fig:L_V_Comparison}) inset is shown the region where the system pressure crosses the saturation line. It is observed in the density profile that the system remains in the liquid phase in this region without presenting a new interface.

The V-L configuration shows a different behavior. The high-temperature region again corresponds to the bottom of the container, but the vapor phase is quite stable in this high temperature phase. This corresponds to the situation shown in Fig.~(\ref{fig:Boiling_Densities}.c).
Only if the gravity was to become so large that the vapor line would contact the top blue line for the vapor instability would we observe pressure induced condensation at the bottom of the container. The amount of gravity required for this is much larger than the values considered in this simulation. Similarly the liquid phase in the low temperature region is stable. We would observe cavitation for very large values of gravity when the liquid line would meet the bottom blue line corresponding to the instability for the liquid. The conditions outlined for the instability of this configuration are very interesting but outside the scope of the current paper. 

\begin{figure}
\centering
\includegraphics[width=\columnwidth]{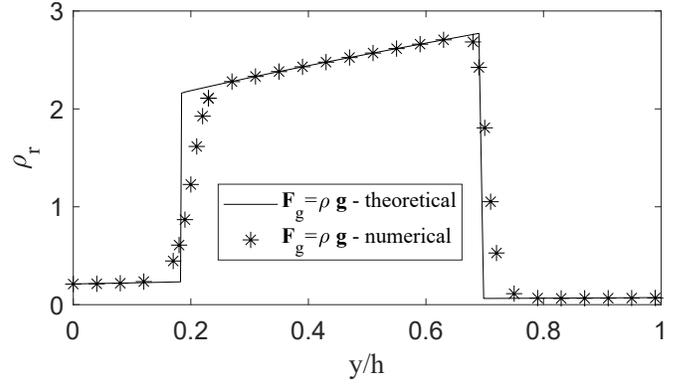}
	\caption{Density profile for the V-L-V example with $g=4\times10^{-5}$ and $\Delta T/T_c=0.3$.}
 \label{fig:V_L_V_Comparison}
\end{figure}

The V-L-V configuration is the most subtle one. This configuration has two interfaces, and both must lie on the saturation pressure line. This severely limits the parameter range for which such morphology can be achieved. For this configuration, when the fluid cross the saturation pressure the fluid switches to a new phase.
As we discussed in the previous paragraphs concerning the L-V configuration, and highlighted in Fig.~(\ref{fig:L_V_Comparison}), the liquid phase can cross the saturation pressure without changing phase. In the case of the V-L-V configuration at this point a transition to a vapor phase will be possible. 
This behavior is shown in Figure~(\ref{fig:V_L_V_Comparison}). 

A fourth and last possible configuration of liquid-vapor-liquid (L-V-L) was not addressed in the present analysis. This last configuration should be much harder to achieve and would require the pressure variation in the vapor to be large enough that two positions on the interfaces can both lie on the saturation pressure line. In this case a large gravity would be required to allow for the vapor line in the $P_r-T_r$ diagram to have a slope larger than the saturation pressure line. This analysis requires much larger gravity than is considered here and is outside the scope of this paper.


\subsection{Configuration Diagrams}
Now that we have considered examples that show that the effect of gravity can allow for stable liquid-vapor coexistence configurations in one dimension, we want to derive more comprehensive state diagrams. 
These diagrams are obtained varying the parameters $g$ and $\Delta T/T_c$ within the ranges $-3\times10^{-4}\leq g\leq3\times10^{-4}$ and $-0.5\leq\Delta T/ T_c\leq 0.5$.
For the two gravity formulations, we evaluated in which regions (in terms of $g$ and $\Delta T/T_c$) each phase configuration could exist.
\begin{figure*}
\centering
	\includegraphics[width=2\columnwidth]{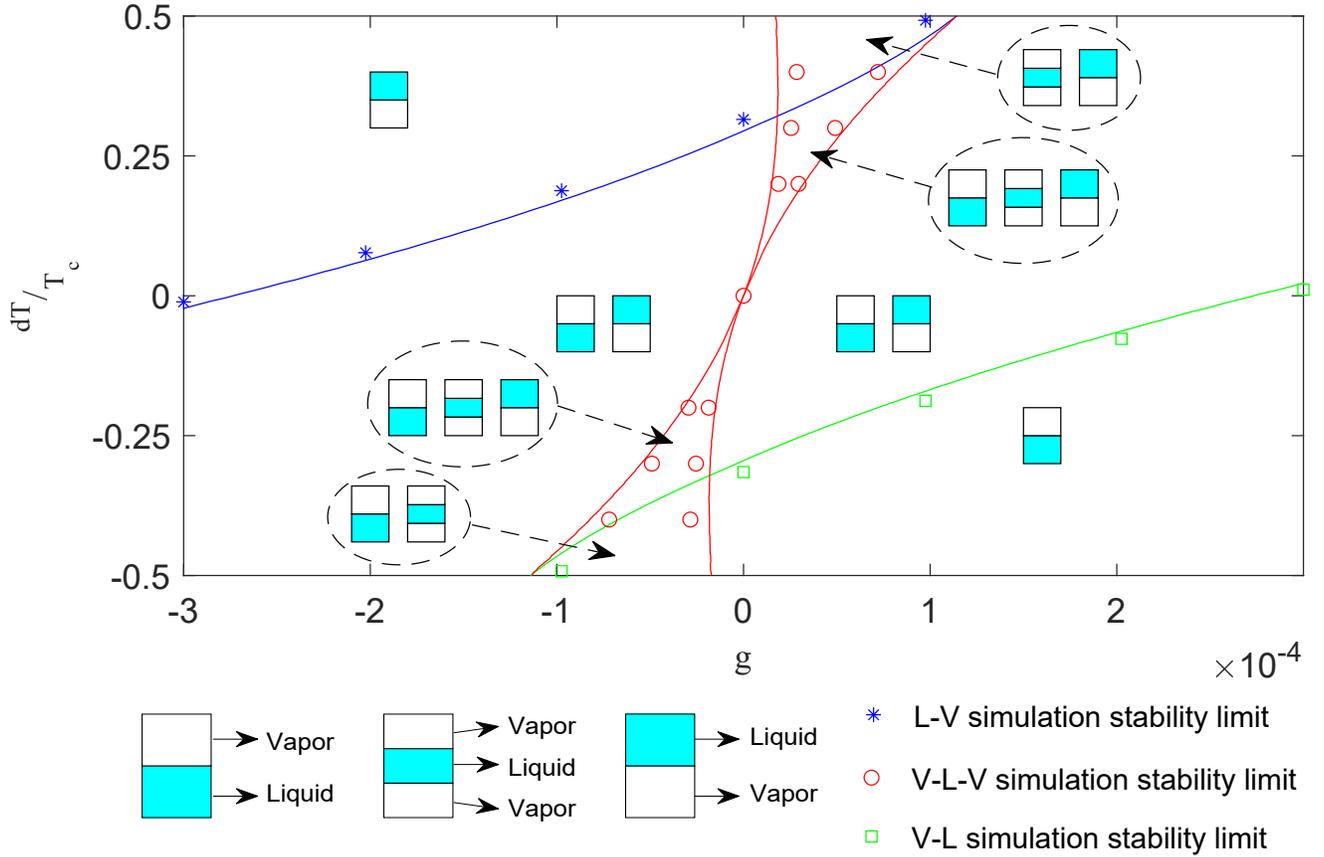}
	\caption{Diagram showing the different phase configurations as a function of $g$ and $\Delta T/T_c$ for the first gravity formulation, Eq~(\ref{eq:Gravity_1}). Solid lines represents the transition regions obtained theoretically.}
	\label{fig:Phase_Diagram_F1}
\end{figure*}
\begin{figure*}
\centering
	\includegraphics[width=2\columnwidth]{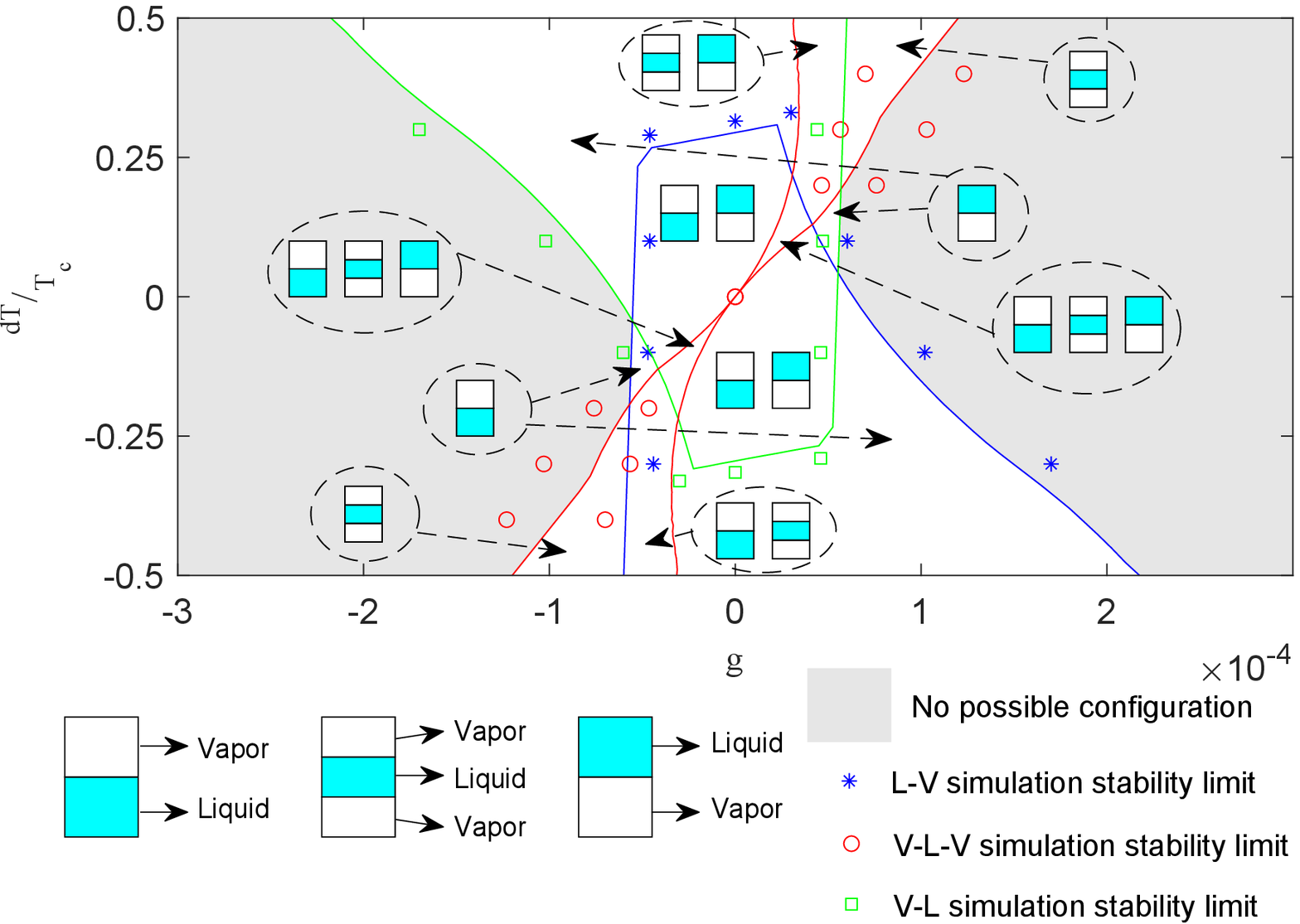}
	\caption{Diagram showing the different phase configurations as a function of $g$ and $\Delta T/T_c$ for the second gravity formulation, Eq~(\ref{eq:Gravity_2}). Solid lines represents the transition regions obtained theoretically.}
	\label{fig:Phase_Diagram_F2}
\end{figure*}

We show the theoretical results and compare them to our lattice Boltzmann simulations.
For each configuration, we started with a lattice Boltzmann simulation known to be stable in terms of $g$ and $\Delta T/T_c$. After that, these parameters were slightly varied until the stability limits of these configurations were found.
It was considered that a limiting case of stability is one in which a variation of only 0.5\% in $g$ or $\Delta T/T_c$ is sufficient for that configuration of phases to cease to exist, either through a phase change process or when the simulation approach zero pressure in the vapor region for the second gravity formulation.

The results obtained are shown in Fig.~(\ref{fig:Phase_Diagram_F1}) and (\ref{fig:Phase_Diagram_F2}). The first figure refers to gravity formulation 1 while the second figure corresponds to formulation 2. The solid lines in the figures correspond to transition regions (obtained theoretically). In these transition regions, on one side of the curve a certain configuration is possible and on the other side of the curve this same configuration can no longer exist. These continuous lines also intersect with each other and in these cases there are regions where more than one configuration is possible. To make the figure interpretation easier, we added small representations of which configurations are possible in each region.

Numerical results of the stability limits of each configuration were represented by small symbols. It can be seen that although there is a small deviation between the numerical and analytical results, in this broad range of gravitational accelerations and temperature differences it is observed a good agreement. As previously mentioned, one of the reasons for these deviations is the wall effect, in which there is a difference between normal pressure and EOS-pressure. Another reason is the fact that the interface is diffuse while in the analytical solution it is considered a discrete interface. As an example, let's consider the V-L-V configuration. In the analytical solution the liquid column can be infinitesimally close to the wall which is not possible in the numerical simulations since the interface has a certain thickness.

The only case in which the deviation between numerical and theoretical results was more significant was for the transition region of the V-L-V configuration with the second gravity formulation in Fig.~(\ref{fig:Phase_Diagram_F2}). 
As shown in Fig.~(\ref{fig:Example_Pressure_Profile}.b), for the second gravity formulation there is a deviation between numerical and theoretical pressure. This occurs because the interface is not at the saturation pressure corresponding to the temperature at that point. For the V-L-V case, it was observed that this deviation is more significant due to the presence of two interfaces. In this way, the theoretical solution does not accurately represent the numerical solution.

Comparing the results of Figs.~(\ref{fig:Phase_Diagram_F1}) and (\ref{fig:Phase_Diagram_F2}) there is a significant difference between the two gravity formulations. Part of these differences had already been mentioned when Fig.~(\ref{fig:Boiling_Theoretical}) was presented, which is embedded in Figs.~(\ref{fig:Phase_Diagram_F1}) and (\ref{fig:Phase_Diagram_F2}). Among these differences between the figures, notice that in Fig.~(\ref{fig:Phase_Diagram_F2}) there is a shaded region. In this region, no configuration is possible, because the system will tend to negative densities, which are not physically possible.

Focusing on the V-L-V configuration in the lower left side of the diagram of Fig.~(\ref{fig:Phase_Diagram_F2}), we observed that this configuration should evolve to L-V configuration in regions situated left to the red line with high negative gravity values. The L-V configuration is unstable and can not exists in this region due to the attainment of zero pressure, and consequently negative densities, in the vapor region. This behavior of L-V configuration was previously explained for the results provided in  Fig.~(\ref{fig:Boiling_Theoretical}). 

Now, focusing on the V-L configuration in the upper left side of the diagram of Fig.~(\ref{fig:Phase_Diagram_F2}), we observed that this configuration should also evolve to L-V configuration in regions situated left to the green line with high negative gravity values. As previously explained, the existence of L-V configuration is not possible in this region. 

Therefore, it can be concluded that in the left shaded sub-region all configurations will tend to the L-V configuration, which becomes unstable and can not exists, as the negative gravity increases in module. Otherwise, in the right shaded sub-region all configurations will tend to the V-L configuration, which will become unstable and not possible for higher positive gravity values. This behavior occurs by the same physical reason related to the attainment of zero pressure (and consequently zero density) in the vapor region.

The reason for the large difference in shape between the two figures (Figs.~(\ref{fig:Phase_Diagram_F1}) and (\ref{fig:Phase_Diagram_F2})) is the large pressure variation in the vapor region for the second formulation. This greater variation induces condensation or negative pressures (and densities) in the vapor region that destabilize the LBM method. This analysis shows that using different gravity formulations can significantly affect LBM results for multiphase problems. For this reason, it is important to be aware of the range of gravitational acceleration values being used.




\section{\label{sec:conclusion}Conclusion}

In this paper we have shown that there can be a significant difference for the results of simulations using the physical force of gravity of $\bm{F}=\rho \bm{g}$ as compared to the more commonly employed force $\bm{F}=(\rho-\langle\rho\rangle)\bm{g}$. While the two formulations give similar results if $g\ll 1$, for more moderate values of gravity typically employed in boiling simulations the differences can be significant. Beyond a certain limit, the second formulation even leads to entirely unphysical regime where no solutions can exist.

In particular we developed a theoretical solution in the sharp interface limit to the phase-coexistence problem and compared this to lattice Boltzmann simulations with a diffuse interface. Both approaches agreed within expected limits. 

We used both approaches to derive a state-diagrams for the possible equilibrium configurations in one dimension for both forms of gravity. These results can be used as a guide that shows for which set of parameters the second gravity formulation can be expected to give physically reasonable results. The advantages of using this second approach, however, are minimal, so we recommend using the physical definition of the force for bounded systems in any case.




\section*{Acknowledgments}

The authors acknowledge the support received from CNPq (National Council for Scientific and Technological Development, process 305941/2020-8) and 
FAPESP (São Paulo Foundation for Research Support, 2016/09509-1 and 2018/09041-5), for developing research that have contributed to this study.




\section*{Data Availability Statement}

The data that support the findings of this study are available from the corresponding author upon reasonable request.




\appendix

\section{Numerical procedure for the theoretical solution}
\label{sec:Appendix_1}

In this appendix we describe the numerical procedure used to solve Eqs.~(\ref{eq:Momentum_Balance_TS}), (\ref{eq:Energy_TS}) and (\ref{eq:Inverse_EOS}). To differentiate this solution from the one obtained by the LBM, 
we call this as theoretical solution. As we already mentioned, it is necessary to know the basic properties $\rho$, $p$ and $T$ at a specific location for both bulk regions. By guessing the interface location $h_i$, the interface temperature $T_i$ and the heat flux $q'$, it is possible to compute the properties at the interface location for both the vapor and liquid phase as described in Eqs.~(\ref{eq:Integral_MaxwellRule}) and (\ref{eq:Pressures_MaxwellRule}). The temperature and pressure varies continuously along the domain and at the interface their values are $p_i$ and $T_i$. At the interface the vapor density is $\rho_{v,i}$ and the liquid density is $\rho_{l,i}$. 

We discretized the system height in a certain number of nodes spaced by a distance $\Delta h$. Then we can obtain the properties at a certain node, for example $p(h+\Delta h)$, as a function of the property in the previous node, $p(h)$, by integrating Eqs.~(\ref{eq:Momentum_Balance_TS}) and (\ref{eq:Energy_TS}) numerically, using an Euler approach with arbitrary accuracy by choosing small values of $\Delta h$.:
\begin{subequations}
\begin{equation}
\begin{aligned}
    \int_{h}^{h+\Delta h}\frac{dp}{dy}dy &= \int_{h}^{h+\Delta h}F_{g,y}(\rho)dy \rightarrow \\
    p(h+\Delta h) - p(h) &= F_{g,y}(\rho(h))\Delta h + O({\Delta h}^2),
\end{aligned}
\end{equation}
\begin{equation}
\begin{aligned}
    \int_{h}^{h+\Delta h}\frac{dT}{dy}dy &= - \int_{h}^{h+\Delta h}\frac{q'}{k}dy \rightarrow \\
    T(h+\Delta h) - T(h) &= -\frac{q'}{k}\Delta h + O({\Delta h}^2).
\end{aligned}
\end{equation}
\end{subequations}
The vapor region is located above the interface and we know the basic properties for this region at the interface $h=h_i$:
\begin{equation} \label{eq:Interface_Vapor}
    \rho_v(h_i) = \rho_{v,i}, ~~ p(h_i) = p_i, ~~ T(h_i) = T_i.
\end{equation}
Then the properties in the rest of the vapor region can be obtained using a simple Euler approach as:
\begin{subequations} \label{eq:Numerical_Vapor_Region}
\begin{equation}
    p(h+\Delta h) = p(h) + F_{g,y}(\rho(h))\Delta h,
\end{equation}
\begin{equation}
    T(h+\Delta h) = T(h) - \frac{q'}{k}\Delta h,
\end{equation}
\begin{equation}
    \rho_v(h+\Delta h) = p_{EOS}^{-1}( p(h+\Delta h), T(h+\Delta h) ).
\end{equation}
\end{subequations}

The same procedure is used in the case of the liquid region:
\begin{equation} \label{eq:Interface_Liquid}
    \rho_l(h_i) = \rho_{l,i}, ~~ p(h_i) = p_i, ~~ T(h_i) = T_i.
\end{equation}
Then the properties in the rest of the vapor region can be obtained by the following equations:
\begin{subequations} \label{eq:Numerical_Liquid_Region}
\begin{equation}
    p(h-\Delta h) = p(h) - F_{g,y}(\rho(h))\Delta h,
\end{equation}
\begin{equation}
    T(h-\Delta h) = T(h) + \frac{q'}{k}\Delta h,
\end{equation}
\begin{equation}
    \rho_l(h-\Delta h) = p_{EOS}^{-1}( p(h-\Delta h), T(h-\Delta h) ).
\end{equation}
\end{subequations}

This approach works well for our purposes, although a higher order discretization could be choosen, if desired.





\providecommand{\noopsort}[1]{}\providecommand{\singleletter}[1]{#1}%

\end{document}